# iNavFIter-M: Matrix Formulation of Functional Iteration for Inertial Navigation Computation

Hongyan Jiang[1], Maoran Zhu[2], Yanyan Fu[3], and Yuanxin Wu[4]

*Shanghai Jiao Tong University, Shanghai, 200240, China*

**The acquisition of attitude, velocity, and position is an essential task in the field of inertial navigation, achieved by integrating the measurements from inertial sensors. Recently, the ultra-precision inertial navigation computation has been tackled by the functional iteration approach (iNavFIter) that drives the non-commutativity errors almost to the computer truncation error level. This paper proposes a computationally efficient matrix formulation of the functional iteration approach, named the iNavFIter-M. The Chebyshev polynomial coefficients in two consecutive iterations are explicitly connected through the matrix formulation, in contrast to the implicit iterative relationship in the original iNavFIter. By so doing, it allows a straightforward algorithmic implementation and a number of matrix factors can be pre-calculated for more efficient computation. Numerical results demonstrate that the proposed iNavFIter-M algorithm is able to achieve the same high computation accuracy as the original iNavFIter does, at the computational cost comparable to the typical two-sample algorithm. The iNavFIter-M algorithm is also implemented on a FPGA board to demonstrate its potential in real time applications.**

---

[1] Doctoral student, Shanghai Key Laboratory of Navigation and Location-based Services, School of Electronic Information and Electrical Engineering, Shanghai Jiao Tong University. AIAA Member.
[2] Doctoral student, Shanghai Key Laboratory of Navigation and Location-based Services, School of Electronic Information and Electrical Engineering, Shanghai Jiao Tong University.
[3] Master student, Shanghai Key Laboratory of Navigation and Location-based Services, School of Electronic Information and Electrical Engineering, Shanghai Jiao Tong University.
[4] Professor, Shanghai Key Laboratory of Navigation and Location-based Services, School of Electronic Information and Electrical Engineering, Shanghai Jiao Tong University.

# I. Introduction

Motion information is essential in many fields, including but not limited to aerospace, driverless vehicles, robotics, and computer vision. Inertial navigation system is a self-contained three-dimensional dead-reckoning navigation system that integrates inertial measurements from accelerometers and gyroscopes to obtain attitude, velocity, and position [1, 2]. With the advances of inertial sensors, the past five decades have witnessed the burst of inertial navigation applications [3, 4]. Even more accurate inertial navigation algorithm will certainly play an important role in sustained high-speed dynamic rigid-body motions [5] and in light of the ultrahigh quality cold-atom inertial sensors under development [6].

The current attitude algorithm structure was established by Jordan [7] and Bortz [8] in early 1970s, which relies on a simplified rotation vector differential equation to be numerically integrated for attitude update [9-11]. Recently, based on the fitted angular velocity polynomial function, a number of works [12-15] have devoted to solving the attitude kinematic equation of different attitude parameterizations, such as the direction cosine matrix (DCM) [16], the rotation vector [9], and the quaternion [14, 15]. In particular, the three-component rotation vector is minimal and does not have any constraint otherwise caused by redundant elements. In view of the polynomial-like attitude differential equation of the three-component Rodrigues vector, the RodFIter method [12, 13] proposed the functional iteration integration approach to solve the attitude kinematic equation and achieved analytical reconstruction of attitude in terms of polynomials. Moreover, the concern about whether the unit-norm constraint of quaternion affects the accuracy has been investigated in [14, 15] and there is no obvious evidence showing that the attitude accuracy is restricted by the quaternion norm.

A subject of research in inertial navigation algorithm constructs devotes to embedding the functions of attitude, velocity, and position within one single generalized framework [5, 17, 18]. For instance, a unified two-speed mathematical framework was described in [17] with new concepts of the so-called velocity and position translation vectors, and the approximate Picard series expansion approach are applied therein. The work [18] tackled the inertial navigation computation problem by using the functional iteration (iNavFIter). It was combined with the Chebyshev polynomial approximation to precisely and efficiently solve the kinematics of attitude, velocity, and position, which is able to reduce the coning/sculling/scrolling errors to almost machine precision. One notable disadvantage of iNavFIter is the computational burden that is about ten times larger than that of the current two-sample algorithm [18].

In fact, the inertial navigation computation is similar in spirit to the initial value problems in the orbit propagation field. In early 1960s, Clenshaw [19] proposed a Picard-Chebyshev method for obtaining the solution to the initial value problems . Feagin [20] presented a matrix formulation of the Picard-Chebyshev method and Fukushima [21] managed to apply the Picard-Chebyshev method to the vector integration of dynamical motions and preliminarily demonstrated its superiority in computational speed. The modified Chebyshev-Picard iteration methods of matrix-vector form was proposed, leading to improved accuracy and significant speedup [22, 23].

Partly motivated by the above-mentioned orbit propagation works, in this paper we try to reformulate the original iNavFIter into a matrix-vector form, named the iNavFIter-M hereafter, for better computational efficiency. The rest of the paper is structured as follows. Section II gives a description of inertial navigation computation problem and a brief review of iNavFIter. In Section III, in terms of the computation of attitude, velocity, and position, the functional iterative integration and Chebyshev polynomial approximation of iNavFIter is recast into the matrix-vector form. Section IV demonstrates, through numerical and FPGA platform assessment, the accuracy and efficiency of the resultant iNavFIter-M algorithm, as compared with the iNavFIter and the typical two-sample algorithm. The conclusions are drawn in Section V

## II. A Brief Review of Inertial Navigation Kinematics and iNavFIter

This section briefly reviews the basic inertial navigation computation problem and the iNavFIter. More details are referred to [18].

### A. Inertial Navigation Kinematics

The Earth-centered inertial frame is denoted by $i$, the Earth-centered Earth-fixed (ECEF) frame is denoted by $e$, the local navigation frame (North-Up-East) is denoted by $n$, and the body frame of the inertial navigation system is denoted by $b$. Choosing an attitude matrix or quaternion to describe the attitude between two coordinate frames, the subscript of the quantity represents the object frame and the superscript represents the reference frame.

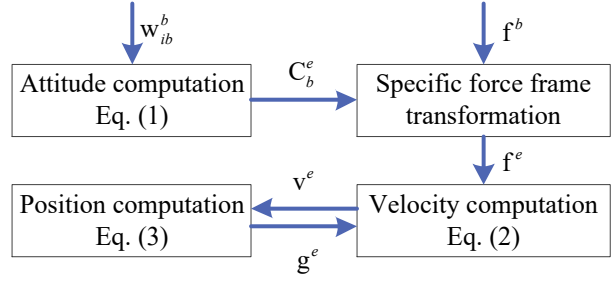

**Fig. 1 Information flow in the ECEF frame mechanization. Arrowed lines indicate information flow directions, and the associated symbols mean that their computation needs to feed on the source information.**

In this paper, the ECEF frame is chosen as the computation reference frame due to its non-singularity [24] and the loosely coupling property among quantities compared with the local navigation frame, as shown in Fig. 1 [18]. In general, the navigation (attitude, velocity, position) rate equations in the ECEF frame are known as [1, 2]

$$\dot{\mathbf{q}}_e^b = \mathbf{q}_e^b \circ \mathbf{w}_{eb}^b / 2 = (\mathbf{q}_e^b \circ \mathbf{w}_{ib}^b - \mathbf{w}_{ie}^e \circ \mathbf{q}_e^b)/2 \tag{1}$$

$$\dot{\mathbf{v}}^e = \mathbf{C}_b^e \mathbf{f}^b - 2\mathbf{w}_{ie}^e \times \mathbf{v}^e + \mathbf{g}^e \tag{2}$$

$$\dot{\mathbf{p}}^e = \mathbf{v}^e \tag{3}$$

where $\mathbf{q}_e^b$ and $\mathbf{C}_e^b$ are the attitude quaternion and attitude matrix, respectively, describing the attitude from the ECEF frame to the body frame, $\mathbf{v}^e = \begin{bmatrix} v_x & v_y & v_z \end{bmatrix}^T$ is the ground velocity expressed in the ECEF frame, $\mathbf{p}^e = \begin{bmatrix} x & y & z \end{bmatrix}^T$ denotes the Cartesian ECEF position, $\mathbf{w}_{eb}^b = \mathbf{w}_{ib}^b - \mathbf{q}_b^e \circ \mathbf{w}_{ie}^e \circ \mathbf{q}_e^b$ is the angular rate of the body frame with respect to the ECEF frame, resolved in the body frame, $\mathbf{w}_{ie}^e = \begin{bmatrix} 0 & 0 & \Omega \end{bmatrix}^T$ is the Earth rotation vector expressed in ECEF frame and $\Omega$ is the magnitude of the earth rotation, $\mathbf{w}_{ib}^b$ and $\mathbf{f}^b$ are the angular rate measured by gyroscopes and the specific force measured by accelerometers, respectively, both expressed in the body frame. A $3\times 3$ skew-symmetric matrix $(\bullet\times)$ is defined to perform the cross product operation for any two three-component vectors, namely, $\mathbf{a}\times\mathbf{b} = (\mathbf{a}\times)\mathbf{b} = [\mathbf{a}]_\times \mathbf{b}$. Besides, $\mathbf{g}^e$ is the gravity vector expressed in the ECEF frame. In general, the gravity models in the literature are described either in the ECEF frame or the local navigation frame, but they are not exactly consistent.

Quaternion in Eq. (1) is a hyper-complex number with four components [25], which can be written as a vector, i.e., $\mathbf{q} = \begin{bmatrix} s & \boldsymbol{\eta}^T \end{bmatrix}^T$, $s$ represents the scalar part and $\boldsymbol{\eta}$ is the three-vector part. The conjugate $\mathbf{q}^*$ of a quaternion is obtained by changing the sign of the three-vector part [26], namely, $\mathbf{q}^* = \begin{bmatrix} s & -\mathbf{\eta}^T \end{bmatrix}^T$. In particular, attitude quaternion used to describe rotation is constrained by unit norm. Treating a three-component vector as a quaternion with zero scalar part, the quaternion and the vector can be multiplied. The symbol $\circ$ in Eq. (1) denotes the product operator of quaternions that is defined as follows [26]

$$\mathbf{q}_1 \circ \mathbf{q}_2 = [\mathbf{q}_1]^+ \begin{bmatrix} s_2 \\ \mathbf{\eta}_2 \end{bmatrix} = [\mathbf{q}_2]^- \begin{bmatrix} s_1 \\ \mathbf{\eta}_1 \end{bmatrix} \tag{4}$$

where $[\mathbf{q}]^+$ and $[\mathbf{q}]^-$ are respectively defined as follows:

$$[\mathbf{q}]^+ = \begin{bmatrix} s & -\mathbf{\eta}^T \\ \mathbf{\eta} & s\mathbf{I}_3 + \mathbf{\eta}\times \end{bmatrix}, \quad [\mathbf{q}]^- = \begin{bmatrix} s & -\mathbf{\eta}^T \\ \mathbf{\eta} & s\mathbf{I}_3 - \mathbf{\eta}\times \end{bmatrix} \tag{5}$$

The quaternion representation of an attitude matrix is given by [27]

$$\mathbf{C}_b^e = (s^2 - \mathbf{\eta}^T\mathbf{\eta})\mathbf{I}_3 + 2\mathbf{\eta}\mathbf{\eta}^T + 2s\mathbf{\eta}\times \tag{6}$$

Consider the navigation updates over the time interval $[0\ t]$, the integrals of Eqs. (1)-(3) yield

$$\mathbf{q} = \mathbf{q}(0) + \int_0^t (\mathbf{q} \circ \mathbf{w}^b - \mathbf{w}^e \circ \mathbf{q})/2\ dt \tag{7}$$

$$\mathbf{v}^e = \mathbf{v}^e(0) + \int_0^t \left( \mathbf{C}_b^e \mathbf{f}^b - 2\mathbf{w}^e \times \mathbf{v}^e + \mathbf{g}^e \right) dt \tag{8}$$

$$\mathbf{p}^e = \mathbf{p}^e(0) + \int_0^t \mathbf{v}^e\ dt \tag{9}$$

where the subscript and superscript in Eqs. (1)-(3) are simplified for the sake of symbolic brevity. $\mathbf{q}(0)$, $\mathbf{v}^e(0)$, and $\mathbf{p}^e(0)$ are the initial values of attitude, velocity, and position, respectively.

## B. The iNavFIter Algorithm

The iNavFIter algorithm is founded on functional iterative integration and Chebyshev polynomials approximation. First of all, the measurements from gyroscopes and accelerometers are fitted by Chebyshev polynomials. The Chebyshev polynomial of the first kind is chosen due to its better numerical stability than the normal polynomial [28].

The Chebyshev polynomial of the first kind is defined over the interval $[-1\ 1]$ through the recurrence relation as

$$F_0(\tau) = 1,\ F_1(\tau) = \tau,\ F_{i+1}(\tau) = 2\tau F_i(\tau) - F_{i-1}(\tau),\ i \geq 1 \tag{10}$$

where $F_i(\tau)$ denotes the Chebyshev polynomial of degree $i$. Noting the Chebyshev polynomial of the first kind is called as Chebyshev polynomial for simplicity. Alternatively, $F_i(\tau)$ can be defined as a trigonometric identity

$$F_i(\tau) = \cos(i \arccos(\tau)), \quad \tau \in [-1\ 1] \tag{11}$$

According to the integral property of Chebyshev polynomial, the following equality can be obtained [28]

$$G_{i,[\tau_{k-1}\ \tau_k]} = \int_{\tau_{k-1}}^{\tau_k} F_i(\tau)d\tau = \begin{cases} \left(\dfrac{iF_{i+1}(\tau_k)}{i^2-1} - \dfrac{\tau_k F_i(\tau_k)}{i-1}\right) - \left(\dfrac{iF_{i+1}(\tau_{k-1})}{i^2-1} - \dfrac{\tau_{k-1}F_i(\tau_{k-1})}{i-1}\right), & i \neq 1 \\ \dfrac{\tau_k^2 - \tau_{k-1}^2}{2}, & i = 1 \end{cases} \tag{12}$$

Besides, for any $j, k \geq 0$, the Chebyshev polynomial satisfies the following equality [28]

$$F_j(\tau)F_k(\tau) = \dfrac{1}{2}\left(F_{j+k}(\tau) + F_{|j-k|}(\tau)\right) \tag{13}$$

In specific, in the light of Eqs. (10), (12), and (13), we have

$$G_{i,[-1\ \tau]} = \int_{-1}^{\tau} F_i(\tau)d\tau = \begin{cases} \left(\dfrac{iF_{i+1}(\tau)}{i^2-1} - \dfrac{\tau F_i(\tau)}{i-1}\right) - \left(\dfrac{iF_{i+1}(-1)}{i^2-1} + \dfrac{F_i(-1)}{i-1}\right)F_0(\tau) \\ = \left(\dfrac{iF_{i+1}(\tau)}{i^2-1} - \dfrac{F_{i+1}(\tau)+F_{|i-1|}(\tau)}{2(i-1)}\right) - \left(\dfrac{iF_{i+1}(-1)}{i^2-1} + \dfrac{F_i(-1)}{i-1}\right)F_0(\tau) \\ = \left(\dfrac{F_{i+1}(\tau)}{2(i+1)} - \dfrac{F_{|i-1|}(\tau)}{2(i-1)}\right) - \dfrac{(-1)^i}{i^2-1}F_0(\tau), & i \neq 1 \\ \dfrac{\tau^2-1}{2} = \dfrac{F_{i+1}(\tau)}{4} - \dfrac{F_0(\tau)}{4}, & i = 1 \end{cases} \tag{14}$$

where $F_i(-1) = (-1)^i$.

At time instants $t_k$ ($k = 1, 2, ..., N$), assuming the discrete angular velocity $\tilde{\mathbf{w}}_{t_k}$ or angular increment $\Delta\tilde{\boldsymbol{\theta}}_{t_k}$ is measured by a triad of gyroscopes and the discrete specific force $\tilde{\mathbf{f}}_{t_k}$ or velocity increments $\Delta\tilde{\mathbf{v}}_{t_k}$ is measured by a triad of accelerometers. For actual time instants $t \in [0\ t_N]$, letting $t = t_N(1+\tau)/2$ to map the time interval onto $[-1\ 1]$. The fitted angular velocity and specific force by the Chebyshev polynomials can be, respectively, written as [12, 18]

$$\mathbf{w}^b = \sum_{i=0}^{n_w} \mathbf{c}_i F_i(\tau), \quad n_w \leq N-1 \tag{15}$$

$$\mathbf{f}^b = \sum_{i=0}^{n_f} \mathbf{d}_i F_i(\tau), \quad n_f \leq N-1 \tag{16}$$

where $n_w$ and $n_f$ are the maximum degrees of Chebyshev polynomials, respectively. Correspondingly, the fitted angular increment and velocity increment can be obtained by

$$\Delta \boldsymbol{\theta}_{t_k} = \int_{t_{k-1}}^{t_k} \mathbf{w}^b \, dt = \frac{t_N}{2} \int_{\tau_{k-1}}^{\tau_k} \mathbf{w}^b \, d\tau = \frac{t_N}{2} \sum_{i=0}^{n_w} \mathbf{c}_i G_{i,[\tau_{k-1}\ \tau_k]} \tag{17}$$

$$\Delta \mathbf{v}_{t_k} = \int_{t_{k-1}}^{t_k} \mathbf{f}^b \, dt = \frac{t_N}{2} \int_{\tau_{k-1}}^{\tau_k} \mathbf{f}^b \, d\tau = \frac{t_N}{2} \sum_{i=0}^{n_f} \mathbf{d}_i G_{i,[\tau_{k-1}\ \tau_k]} \tag{18}$$

Then the coefficients $\mathbf{c}_i$ and $\mathbf{d}_i$ could be determined by solving the least-square equations of discrete measurements from gyroscopes and accelerometers [12, 18].

Hereafter, the functional iteration technique is applied to the integrations of attitude, velocity, and position, respectively. Firstly, for attitude computation, the attitude quaternion in Eq. (7) can be iteratively computed by

$$\begin{aligned} \mathbf{q}_{l+1} &= \mathbf{q}(0) + \frac{1}{2} \int_0^t \left[ \mathbf{q}_l(t) \circ \mathbf{w}^b(t) - \mathbf{w}^e \circ \mathbf{q}_l(t) \right] dt \\ &= \mathbf{q}(0) + \frac{t_N}{4} \int_{-1}^{\tau} \left[ \mathbf{q}_l(\tau) \circ \mathbf{w}^b(\tau) - \mathbf{w}^e \circ \mathbf{q}_l(\tau) \right] d\tau \end{aligned} \tag{19}$$

where $\mathbf{q}_0(t) \equiv \mathbf{q}(0)$ is chosen as the initial attitude quaternion function, and the subscript $l$ denotes the $l$th iteration. If the quaternion estimator in the $l$th iteration is approximated by Chebyshev polynomials, namely,

$$\mathbf{q}_l = \sum_{i=0}^{m_q} \mathbf{b}_{l,i} F_i(\tau) \tag{20}$$

where $m_q$ is the prescribed maximum degree of Chebyshev polynomial, and $\mathbf{b}_{l,i}$ denotes the coefficient of Chebyshev polynomial of $i$th degree in the $l$th iteration. The quaternion in the next iteration could be obtained as follows [18]:

$$\begin{aligned} \mathbf{q}_{l+1} &= \mathbf{q}(0) + \frac{t_N}{8} \left( \sum_{i=0}^{m_q} \sum_{j=0}^{n_w} \mathbf{b}_{l,i} \circ \mathbf{c}_j \left( G_{i+j,[-1\ \tau]} + G_{|i-j|,[-1\ \tau]} \right) - 2\sum_{i=0}^{m_q} \boldsymbol{\omega}_e \circ \mathbf{b}_{l,i} G_{i,[-1\ \tau]} \right) \\ &= \sum_{i=0}^{m_q+n_w+1} \mathbf{b}_{l+1,i} F_i(\tau) \quad \overset{\text{polynomial truncation}}{\approx} \quad \sum_{i=0}^{m_q} \mathbf{b}_{l+1,i} F_i(\tau) \end{aligned} \tag{21}$$

Note that, the last approximation accounts for the polynomial truncation with a prescribed maximum degree $m_q$ [13].

Similarly, the velocity and position in Eqs. (8) and (9), respectively, can be iteratively computed by

$$\begin{aligned}
\mathbf{v}_{l+1}^e &= \mathbf{v}^e(0) + \int_0^t \left[ \mathbf{C}_b^e(t)\mathbf{f}^b(t) - 2\mathbf{w}^e \times \mathbf{v}_l^e(t) + \mathbf{g}^e\left(\mathbf{p}_l^e(t)\right) \right] dt \\
&= \mathbf{v}^e(0) + \frac{t_N}{2} \int_{-1}^\tau \left[ \mathbf{C}_b^e(\tau)\mathbf{f}^b(\tau) - 2\mathbf{w}^e \times \mathbf{v}_l^e(\tau) + \mathbf{g}^e\left(\mathbf{p}_l^e(\tau)\right) \right] d\tau
\end{aligned} \quad (22)$$

$$\mathbf{p}_{l+1}^e = \mathbf{p}^e(0) + \int_0^t \mathbf{v}_l^e(t)\, dt = \mathbf{p}^e(0) + \frac{t_N}{2} \int_{-1}^\tau \mathbf{v}_l^e(\tau)\, d\tau \quad (23)$$

where the initial velocity/position functions are chosen as $\mathbf{v}_0^e(t) \equiv \mathbf{v}^e(0)$ and $\mathbf{p}_0^e(t) \equiv \mathbf{p}^e(0)$. Moreover, the velocity, position, as well as the gravity can also be approximated by Chebyshev polynomials, say

$$\mathbf{v}_l^e = \sum_{i=0}^{m_v} \mathbf{s}_{l,i} F_i(\tau) \quad (24)$$

$$\mathbf{p}_l^e = \sum_{i=0}^{m_p} \boldsymbol{\rho}_{l,i} F_i(\tau) \quad (25)$$

$$\mathbf{g}^e\left(\mathbf{p}_l^e\right) = \sum_{i=0}^{m_g} \boldsymbol{\gamma}_{l,i} F_i(\tau) \quad (26)$$

with $m_v, m_p, m_g$ the maximum degrees of Chebyshev polynomial and $\mathbf{s}_{l,i}$, $\boldsymbol{\rho}_{l,i}$ the coefficients of Chebyshev polynomial of the *i*th degree in the *l*th iteration. Note that, as for the gravity function described by Chebyshev polynomials, the coefficients $\boldsymbol{\gamma}_{l,i}$ cannot be obtained from the previous iteration. According to the orthogonal property of Chebyshev polynomials, the Chebyshev polynomial coefficients of gravity are given by [18, 28, 29]

$$\begin{aligned}
\boldsymbol{\gamma}_{l,i} &\approx \frac{2-\delta_{0i}}{m_g+1} \sum_{k=0}^{m_g} \cos\left(\frac{i(k+1/2)\pi}{m_g+1}\right) \mathbf{g}^e\left(\mathbf{p}_l^e\left(\cos\left(\frac{(k+1/2)\pi}{m_g+1}\right)\right)\right) \\
&= \frac{2-\delta_{0i}}{m_g+1} \sum_{k=0}^{m_g} \mathbf{g}^e\left(\mathbf{p}_l^e(\sigma_k)\right) F_i(\sigma_k)
\end{aligned} \quad (27)$$

where $\sigma_k$ denote the Chebyshev-roots points [29], defined by

$$\sigma_k = \cos\left(\frac{(k+1/2)\pi}{m_g+1}\right), \quad k = 0,1,\cdots,m_g \quad (28)$$

and $\delta_{0i}$ is the Kronecker delta function, yielding 1 for $i=0$ and zero otherwise. The exact coefficients in Eq. (27) could be obtained only if the number of summation terms, $m_g+1$, approaches infinity

Since then, the velocity and position in the next iteration could be obtained by [18]

$$\mathbf{v}_{l+1}^e = \mathbf{v}^e(0) + \frac{t_N}{2}\left(\mathbf{I}_f - 2\sum_{i=0}^{m_v}\boldsymbol{\omega}_e \times \mathbf{s}_{l,i}G_{i,[-1\ \tau]} + \sum_{i=0}^{m_g}\boldsymbol{\gamma}_{l,i}G_{i,[-1\ \tau]}\right)$$

$$= \sum_{i=0}^{\max\{2m_q+n_f,m_v,m_g\}+1}\mathbf{s}_{l+1,i}F_i(\tau) \stackrel{\text{polynomial truncation}}{\approx} \sum_{i=0}^{m_v}\mathbf{s}_{l+1,i}F_i(\tau) \tag{29}$$

$$\mathbf{p}_{l+1}^e = \mathbf{p}(0) + \frac{t_N}{2}\sum_{i=0}^{m_v}\mathbf{s}_{l,i}G_{i,[-1\ \tau]} = \sum_{i=0}^{m_v+1}\boldsymbol{\rho}_{l+1,i}F_i(\tau) \stackrel{\text{polynomial truncation}}{\approx} \sum_{i=0}^{m_p}\boldsymbol{\rho}_{l+1,i}F_i(\tau) \tag{30}$$

where $\mathbf{I}_f$ is defined by

$$\mathbf{I}_f(\tau) = \int_{-1}^{\tau}\mathbf{C}_b^e\mathbf{f}^b d\tau = \int_{-1}^{\tau}\mathbf{q}_{l+1} \circ \mathbf{f}^b \circ \mathbf{q}_{l+1}^* d\tau$$

$$= \int_{-1}^{\tau}\left(\sum_{i=0}^{m_q}\mathbf{b}_{l+1,i}F_i(\tau)\right) \circ \left(\sum_{j=0}^{n_f}\mathbf{d}_j F_j(\tau)\right) \circ \left(\sum_{k=0}^{m_q}\mathbf{b}_{l+1,k}^* F_k(\tau)\right)d\tau \tag{31}$$

$$= \frac{1}{4}\sum_{i=0}^{m_q}\sum_{j=0}^{n_f}\sum_{k=0}^{m_q}\mathbf{b}_{l+1,i} \circ \mathbf{d}_j \circ \mathbf{b}_{l+1,k}^* \begin{pmatrix} G_{i+j+k,[-1\ \tau]} + G_{|i+j-k|,[-1\ \tau]} \\ + G_{|i-j|+k,[-1\ \tau]} + G_{||i-j|-k|,[-1\ \tau]} \end{pmatrix}$$

The technique of polynomial truncation is also applied to both velocity and position computation in the last step. It is important to note that the velocity and position are iteratively computed together due to their inherently mutual dependence on each other. Besides, the iteration process of velocity and position will not be started until the attitude iteration has been finished.

To sum up, after two consecutive processes, namely, the attitude iterative computation as well as the velocity/position iterative computation, the attitude, velocity and position could be exactly obtained. Numerical tests in [18] show the noncommutativity errors could be significantly reduced to the level of machine precision, having remarkable accuracy superiority over the typical two-sample algorithm. However, in terms of computational efficiency, the iNavFIter is inferior owing to the complexity of iterative process. It is unfriendly to real time implementations. It may be argued that the computation of iNavFIter could be optimized. To be specific, take the attitude quaternion iteration in Eq. (21) as an example, 16 scalar multiplications are included in a quaternion multiplication and the computational complexity is roughly proportional to $O(16(m_q+1)(n_w+1)+16(m_q+1))$ [18]. It will be proportional to $O(1400)$ for $m_q = 9$ and $n_w = 7$ at each iteration. Note that the last step in Eq. (21) is to truncate the polynomial terms higher than $m_q$. If the computation of those terms were spared in some way, the computational complexity could be reduced to $O(16 \times 2(m_q+1)) = O(320)$ at each iteration, which means a complexity reduction of about five times. In next section, the iterative process of iNavFIter will be formulated explicitly in a matrix form so as to improve the computational efficiency.

## III. iNavFIter-M: iNavFIter in Matrix Form

### A. Attitude Computation

The functional iteration of attitude quaternion in Eq. (19) can be rewritten as

$$\mathbf{q}_{l+1}(\tau) = \mathbf{q}(0) + \frac{t_N}{4} \int_{-1}^{\tau} \mathbf{r}(\tau, \mathbf{q}_l(\tau)) d\tau \tag{32}$$

where the integrand function $\mathbf{r}(\tau, \mathbf{q}_l(\tau))$ is obviously defined as

$$\mathbf{r}(\tau, \mathbf{q}_l(\tau)) = \mathbf{q}_l(\tau) \circ \mathbf{w}^b(\tau) - \mathbf{w}^e \circ \mathbf{q}_l(\tau) \tag{33}$$

If $\mathbf{r}(\tau, \mathbf{q}_l(\tau))$ is a Lipschitz continuous function, it can be approximated by Chebyshev polynomials as follows

$$\mathbf{r}(\tau, \mathbf{q}_l(\tau)) \approx \sum_{i=0}^{m_q} \boldsymbol{\alpha}_{l,i} F_i(\tau) \tag{34}$$

In the same way as the calculation of Chebyshev coefficient of gravity function in Eq. (27), the coefficient $\boldsymbol{\alpha}_{l,i}$ can be obtained by

$$\boldsymbol{\alpha}_{l,i} = \frac{2 - \delta_{0i}}{m_q + 1} \sum_{k=0}^{m_q} \mathbf{r}(\sigma_k, \mathbf{q}_l(\sigma_k)) F_i(\sigma_k) \tag{35}$$

Equations (34) and (35) indicate that the coefficient $\boldsymbol{\alpha}_{l,i}$ and the function value $\mathbf{r}(\tau, \mathbf{q}_l(\tau))$ can be changed to each other conveniently, in other words, they are computationally equivalent. Note that the relationship between Chebyshev coefficient of integrand function $\boldsymbol{\alpha}_{l,i}$ and attitude quaternion value at Chebyshev-root point $\mathbf{q}_l(\sigma_k)$ is essential in the subsequent development to obtain a compact matrix form. We define

$$\boldsymbol{\alpha}_l = [\boldsymbol{\alpha}_{l,0}, \boldsymbol{\alpha}_{l,1}, \cdots, \boldsymbol{\alpha}_{l,m_q}]^T \tag{36}$$

$$\mathbf{R} = \left[ \mathbf{r}(\sigma_0, \mathbf{q}_l(\sigma_0)), \cdots, \mathbf{r}(\sigma_{m_q}, \mathbf{q}_l(\sigma_{m_q})) \right]^T \triangleq \left[ \mathbf{r}(\mathbf{q}_{l,0}), \cdots, \mathbf{r}(\mathbf{q}_{l,m_q}) \right]^T \tag{37}$$

According to Eq. (36), the following matrix identity can be obtained

$$\boldsymbol{\alpha}_l = \begin{bmatrix} \boldsymbol{\alpha}_{l,0}, & \boldsymbol{\alpha}_{l,1}, & \cdots, & \boldsymbol{\alpha}_{l,m_q} \end{bmatrix}^T$$

$$= \frac{2}{m_q+1} \begin{bmatrix} \frac{1}{2}\mathbf{r}^T(\mathbf{q}_{l,0})F_0(\sigma_0) + \cdots + \frac{1}{2}\mathbf{r}^T(\mathbf{q}_{l,m_q})F_0(\sigma_{m_q}) \\ \mathbf{r}^T(\mathbf{q}_{l,0})F_1(\sigma_0) + \cdots + \mathbf{r}^T(\mathbf{q}_{l,m_q})F_1(\sigma_{m_q}) \\ \vdots \\ \mathbf{r}^T(\mathbf{q}_{l,0})F_{m_q}(\sigma_0) + \cdots + \mathbf{r}^T(\mathbf{q}_{l,m_q})F_{m_q}(\sigma_{m_q}) \end{bmatrix} \quad (38)$$

$$\triangleq \frac{2}{m_q+1}\mathbf{Z}\mathbf{F}^T\mathbf{R}$$

with the definition

$$\mathbf{Z} = diag\left([1/2,\ 1,\ \cdots,\ 1]\right) \quad (39)$$

$$\mathbf{F} = \begin{bmatrix} F_0(\sigma_0) & F_1(\sigma_0) & \cdots & F_M(\sigma_0) \\ F_0(\sigma_1) & F_1(\sigma_1) & \cdots & F_M(\sigma_1) \\ \vdots & \vdots & & \vdots \\ F_0(\sigma_M) & F_1(\sigma_M) & \cdots & F_M(\sigma_M) \end{bmatrix} \quad (40)$$

and $M$ is the maximum degree of Chebyshev polynomials and is set to $m_q$ for attitude computation. Once the parameter $M$ is prescribed, matrices $\mathbf{Z}$ and $\mathbf{F}$ would both be constant. In terms of the calculation of matrix $\mathbf{R}$, from Eqs. (33) and (37), we have

$$\mathbf{R} = \begin{bmatrix} \mathbf{r}(\mathbf{q}_{l,0}), & \cdots, & \mathbf{r}(\mathbf{q}_{l,m_q}) \end{bmatrix}^T$$

$$= \begin{bmatrix} \left(\mathbf{q}_l(\sigma_0) \circ \mathbf{w}^b(\sigma_0) - \mathbf{w}^e \circ \mathbf{q}_l(\sigma_0)\right)^T \\ \vdots \\ \left(\mathbf{q}_l(\sigma_{m_q}) \circ \mathbf{w}^b(\sigma_{m_q}) - \mathbf{w}^e \circ \mathbf{q}_l(\sigma_{m_q})\right)^T \end{bmatrix} = \begin{bmatrix} \left[\left(\left[\mathbf{w}^b(\sigma_0)\right]^- - \left[\mathbf{w}^e\right]^+\right)\mathbf{q}_l(\sigma_0)\right]^T \\ \vdots \\ \left[\left(\left[\mathbf{w}^b(\sigma_{m_q})\right]^- - \left[\mathbf{w}^e\right]^+\right)\mathbf{q}_l(\sigma_{m_q})\right]^T \end{bmatrix} \quad (41)$$

$$\triangleq diag(\mathbf{Q}_l)\left(\mathbf{W}^{b-} - \mathbf{W}^{e+}\right)$$

For a $n$ by $m$ matrix $\mathbf{A} = [\boldsymbol{a}_1,\ \boldsymbol{a}_2,\ \cdots,\ \boldsymbol{a}_n]^T$, where $\boldsymbol{a}_i$, $i=1,\cdots,n$ is a m-dimensional vector, we have

$$diag(\mathbf{A}) = \begin{bmatrix} \boldsymbol{a}_1^T & 0 & \cdots & 0 \\ 0 & \boldsymbol{a}_2^T & 0 & \vdots \\ \vdots & 0 & \ddots & 0 \\ 0 & 0 & \cdots & \boldsymbol{a}_n^T \end{bmatrix} \quad (42)$$

In Eq. (41), we define

$$\mathbf{Q}_l = \begin{bmatrix} \mathbf{q}_l(\sigma_0), & \mathbf{q}_l(\sigma_1), & \cdots, & \mathbf{q}_l(\sigma_{m_q}) \end{bmatrix}^T$$

$$\mathbf{W}^{b-} = \begin{bmatrix} \left[\overline{\mathbf{w}^b(\sigma_0)}\right], \left[\overline{\mathbf{w}^b(\sigma_1)}\right], \cdots, \left[\overline{\mathbf{w}^b(\sigma_{m_q})}\right] \end{bmatrix}^T \quad (43)$$

$$\mathbf{W}^{e+} = \begin{bmatrix} \left[\overset{+}{\mathbf{w}^e}\right], & \left[\overset{+}{\mathbf{w}^e}\right], & \cdots, & \left[\overset{+}{\mathbf{w}^e}\right] \end{bmatrix}^T$$

where $\mathbf{q}_l(\sigma_k)$, $k=0,1,...,m_q$ are the attitude quaternions at Chebyshev-roots points, $\mathbf{w}^b(\sigma_k)$, $k=0,1,...,m_q$ can be readily obtained from Eq. (15) or Eq. (17), $\mathbf{w}^e$ is the Earth rotation vector. Note that the function $\mathbf{r}(\mathbf{q}_{l,k})$, $k=0,\cdots,m_q$ can be calculated in a parallel mechanism since the function values at different Chebyshev-roots points are independent of each other. It would play an important role in algorithm speed enhancement, especially for cases where the evaluation of function values costs a lot of computations.

In addition, the attitude quaternion in the current iteration could also be approximated by Chebyshev polynomials with maximum degree $m_q$

$$\mathbf{q}_{l+1}(\tau) = \sum_{i=0}^{m_q} \mathbf{b}_{l+1,i} F_i(\tau) \quad (44)$$

Therefore, using Eq. (32), (34), and (44), we have

$$\begin{aligned}
&\sum_{i=0}^{m_q} \mathbf{b}_{l+1,i} F_i(\tau) \\
&\approx \mathbf{q}(0) + \frac{t_N}{4} \int_{-1}^{\tau} \sum_{i=0}^{m_q} \mathbf{a}_{l,i} F_i(\tau)\, d\tau \\
&= \mathbf{q}(0) + \frac{t_N}{4} \mathbf{a}_{l,0} F_1(\tau) + \frac{t_N}{4} \mathbf{a}_{l,0} + \frac{t_N}{4} \frac{\mathbf{a}_{l,1}}{4} F_2(\tau) - \frac{t_N}{4} \frac{\mathbf{a}_{l,1}}{4} + \frac{t_N}{4} \sum_{i=2}^{m_q} \frac{\mathbf{a}_{l,i}}{2} \left( \frac{F_{i+1}(\tau)}{i+1} - \frac{(-1)^{i+1}}{i+1} - \frac{F_{i-1}(\tau)}{i-1} + \frac{(-1)^{i+1}}{i-1} \right) \\
&= \mathbf{q}(0) + \frac{t_N}{4} \mathbf{a}_{l,0} - \frac{t_N}{16} \mathbf{a}_{l,1} + \frac{t_N}{4} \sum_{i=2}^{m_q} \mathbf{a}_{l,i} \frac{(-1)^{i+1}}{i^2-1} + \frac{t_N}{4} \mathbf{a}_{l,0} F_1(\tau) + \frac{t_N}{4} \frac{\mathbf{a}_{l,1}}{4} F_2(\tau) + \frac{t_N}{4} \sum_{i=2}^{m_q} \frac{\mathbf{a}_{l,i}}{2} \left( \frac{F_{i+1}(\tau)}{i+1} - \frac{F_{i-1}(\tau)}{i-1} \right) \\
&= \mathbf{\kappa} + \frac{t_N}{4} \left( \mathbf{a}_{l,0} - \frac{\mathbf{a}_{l,2}}{2} \right) F_1(\tau) + \frac{t_N}{8} \sum_{i=2}^{m_q-1} \frac{\mathbf{a}_{l,i-1} - \mathbf{a}_{l,i+1}}{i} F_i(\tau) + \frac{t_N}{8} \frac{\mathbf{a}_{l,m_q-1}}{m_q} F_{m_q}(\tau) + \frac{t_N}{8} \frac{\mathbf{a}_{l,m_q}}{m_q+1} F_{m_q+1}(\tau)
\end{aligned} \quad (45)$$

where the property of Chebyshev polynomials in Eq. (14) is used therein, and $\mathbf{\kappa}$ is defined as follows

$$\mathbf{\kappa} = \mathbf{q}(0) + \frac{t_N}{4} \mathbf{a}_{l,0} - \frac{t_N}{16} \mathbf{a}_{l,1} + \frac{t_N}{4} \sum_{i=2}^{m_q} \mathbf{a}_{l,i} \frac{(-1)^{i+1}}{i^2-1} \quad (46)$$

Comparing both sides of Eq. (45), the following identity can be obtained

$$\begin{cases} \mathbf{b}_{l+1,0} = \boldsymbol{\kappa}, \\ \mathbf{b}_{l+1,1} = \dfrac{t_N}{4}\left(\boldsymbol{\alpha}_{l,0} - \dfrac{\boldsymbol{\alpha}_{l,2}}{2}\right), \\ \mathbf{b}_{l+1,i} = \dfrac{t_N}{8}\dfrac{\boldsymbol{\alpha}_{l,i-1} - \boldsymbol{\alpha}_{l,i+1}}{i}, \quad i = 2,\cdots,m_q - 1 \\ \mathbf{b}_{l+1,m_q} = \dfrac{t_N}{8}\dfrac{\boldsymbol{\alpha}_{l,m_q-1}}{m_q}, \end{cases} \qquad (47)$$

It can be found that a formula relationship between the Chebyshev coefficients of integrand function and the Chebyshev coefficients of attitude quaternion function in two consecutive iterations is established.

The coefficients of Chebyshev polynomials can be put into a matrix, defined as follows

$$\mathbf{b}_{l+1} = \left[\mathbf{b}_{l+1,0}, \mathbf{b}_{l+1,1}, \cdots, \mathbf{b}_{\boldsymbol{\alpha}_{l+1,m_q}}\right]^T \qquad (48)$$

Therefore, in the light of Eq. (38), (41), (46) and (47), the following matrix identity can be obtained

$$\begin{aligned}
\mathbf{b}_{l+1} &= \left[\mathbf{b}_{l+1,0},\ \mathbf{b}_{l+1,1},\ \mathbf{b}_{l+1,2},\ \cdots,\ \mathbf{b}_{l+1,r},\ \cdots,\ \mathbf{b}_{l+1,m_q}\right]^T \\
&= \begin{bmatrix} \mathbf{q}^T(0) + \dfrac{t_N}{4}\boldsymbol{\alpha}_{l,0}^T - \dfrac{t_N}{16}\boldsymbol{\alpha}_{l,1}^T + \dfrac{t_N}{4}\displaystyle\sum_{i=2}^{m_q}\boldsymbol{\alpha}_{l,i}^T \dfrac{(-1)^{i+1}}{i^2-1} \\ \dfrac{t_N}{4}\dfrac{1}{2\times 1}(2\boldsymbol{\alpha}_{l,0}^T - \boldsymbol{\alpha}_{l,2}^T) \\ \dfrac{t_N}{4}\dfrac{1}{2\times 2}(\boldsymbol{\alpha}_{l,1}^T - \boldsymbol{\alpha}_{l,3}^T) \\ \vdots \\ \dfrac{t_N}{4}\dfrac{1}{2\times r}(\boldsymbol{\alpha}_{l,r-1}^T - \boldsymbol{\alpha}_{l,r+1}^T) \\ \vdots \\ \dfrac{t_N}{4}\dfrac{\boldsymbol{\alpha}_{l,m_q-1}^T}{2m_q} \end{bmatrix} \\
&\triangleq \boldsymbol{\chi}_0 + \dfrac{t_N}{4}\mathbf{UD}\boldsymbol{\alpha}_l = \boldsymbol{\chi}_0 + \dfrac{t_N}{4}\cdot\dfrac{2}{m_q+1}\mathbf{UDZF}^T\mathbf{R} \\
&= \boldsymbol{\chi}_0 + \dfrac{t_N}{2(m_q+1)}\mathbf{C}_s\cdot diag(\mathbf{Q}_l)\left(\mathbf{W}^{b-} - \mathbf{W}^{e+}\right)
\end{aligned} \qquad (49)$$

with the definition

$$\boldsymbol{\chi}_0 = \left[\mathbf{q}(0),\ 0,\ \cdots,\ 0\right]^T \qquad (50)$$

$$\mathbf{C}_s = \mathbf{UDZF}^T \qquad (51)$$

$$\mathbf{U} = diag\left(\left[1, \frac{1}{2}, \frac{1}{4}, \cdots, \frac{1}{2(M-1)}, \frac{1}{2M}\right]\right) \tag{52}$$

$$\mathbf{D} = \begin{bmatrix} 1 & -\frac{1}{4} & -\frac{1}{3} & \frac{1}{8} & -\frac{1}{15} & \cdots & \frac{(-1)^{M+1}}{M^2-1} \\ 2 & 0 & -1 & 0 & 0 & \cdots & 0 \\ 0 & 1 & 0 & -1 & 0 & \cdots & 0 \\ 0 & 0 & 1 & 0 & -1 & \cdots & 0 \\ \vdots & \vdots & \vdots & \vdots & \vdots & \cdots & \vdots \\ 0 & 0 & 0 & 0 & 0 & \cdots & 1 & 0 & -1 \\ 0 & 0 & 0 & 0 & 0 & \cdots & 0 & 1 & 0 \end{bmatrix} \tag{53}$$

where we set $M = m_q$, and the element in the $j$th ($3 \leq j \leq M+1$) column and the first row of matrix $\mathbf{D}$ is given by $\mathbf{D}[1, j] = (-1)^j / ((j-1)^2 - 1)$. $\mathbf{C}_s$ will be a constant matrix once the maximum degree of Chebyshev polynomials is prescribed, so it can be computed beforehand and saved in the computer memory. This will significantly improve the computational efficiency and do not have to be computed in real time in contrast to the computation process of iNavFIter [18]. Note that, the matrix $\mathbf{R}$ is used for the calculation of the polynomial coefficient matrix $\mathbf{b}_{l+1}$, in which the attitude quaternion values at Chebyshev-roots points in the last iteration, i.e., $\mathbf{Q}_l$, are required.

According to Eq. (44), the following matrix identity can be obtained

$$\begin{aligned}
\mathbf{Q}_{l+1} &= \left[\mathbf{q}_{l+1}(\sigma_0), \cdots, \mathbf{q}_{l+1}(\sigma_{m_q})\right]^T \\
&= \begin{bmatrix} \mathbf{b}_{l+1,0}^T F_0(\sigma_0) + \cdots + \mathbf{b}_{l+1,m_q}^T F_{m_q}(\sigma_0) \\ \vdots \\ \mathbf{b}_{l+1,0}^T F_0(\sigma_{m_q}) + \cdots + \mathbf{b}_{l+1,m_q}^T F_{m_q}(\sigma_{m_q}) \end{bmatrix} = \begin{bmatrix} F_0(\sigma_0) & \cdots & F_{m_q}(\sigma_0) \\ \vdots & \vdots & \vdots \\ F_0(\sigma_{m_q}) & \cdots & F_{m_q}(\sigma_{m_q}) \end{bmatrix} \begin{bmatrix} \mathbf{b}_{l+1,0}^T \\ \vdots \\ \mathbf{b}_{l+1,m_q}^T \end{bmatrix} \\
&= \mathbf{F}\mathbf{b}_{l+1}
\end{aligned} \tag{54}$$

where the matrix $\mathbf{F}$ is defined as Eq. (40), and $M$ is set to $m_q$. In this way, the values of attitude quaternion at Chebyshev-roots points are updated with the help of coefficient matrix $\mathbf{b}_{l+1}$, which has already been obtained using Eq. (49). Subsequently, these new attitude quaternion values at Chebyshev-roots points will be used to calculate the coefficient matrix for the next new iteration.

It can be seen that both the new proposed method and the iNavFIter [18] method depend on the update of coefficients to achieve the functional iteration. The difference is that the new proposed method updates the coefficients with the help of attitude quaternion values at Chebyshev-roots points (as seen in Eqs. (32)-(35)), so that the iterative process can be expressed in a matrix formulation of finite dimension, while the iNavFIter utilizes the direct expansion of Eq. (32) and spends significant computation resources on high-order terms.

The Chebyshev coefficients of attitude quaternion function will be iteratively computed until a prescribed accuracy criterion is satisfied or a maximum iteration number is met. For instance, the root mean square (RMS) error could be considered as the accuracy criterion, defined as

$$\mathbf{e}_q = \sqrt{\sum_{i=0}^{m_q}\left(\mathbf{b}_{l+1,i}-\mathbf{b}_{l,i}\right)^T \left(\mathbf{b}_{l+1,i}-\mathbf{b}_{l,i}\right) \Big/ \left(m_q+1\right)} \tag{55}$$

The attitude quaternion computed by the resultant Chebyshev coefficients according to Eq. (44) is denoted by $\mathbf{q}(\tau)$.

## B. Velocity Computation

The velocity in the ECEF frame can be iteratively computed in the same way. The functional iteration of velocity in Eq. (22) can be represented as

$$\mathbf{v}_{l+1}^e(\tau) = \mathbf{v}^e(0) + \frac{t_N}{2}\int_{-1}^{\tau}\mathbf{y}(\tau,\mathbf{v}_l^e(\tau))\,d\tau \tag{56}$$

where the integrand function $\mathbf{y}(\tau,\mathbf{v}_l^e(\tau))$ is defined as

$$\begin{aligned}\mathbf{y}(\tau,\mathbf{v}_l^e(\tau)) &= \mathbf{C}_b^e(\tau)\mathbf{f}^b(\tau) - 2\mathbf{w}^e \times \mathbf{v}_l^e(\tau) + \mathbf{g}^e\left(\mathbf{p}_l^e(\tau)\right) \\ &= \mathbf{q}(\tau)\circ\mathbf{f}^b(\tau)\circ\mathbf{q}^*(\tau) - 2\mathbf{w}^e\times\mathbf{v}_l^e(\tau) + \mathbf{g}^e(\mathbf{p}_l^e(\tau))\end{aligned} \tag{57}$$

Similarly, the integrand $\mathbf{y}(\tau,\mathbf{v}_l^e(\tau))$ can be approximated by Chebyshev polynomials with the maximum degree of $m_v$, that is to say,

$$\mathbf{y}(\tau,\mathbf{v}_l^e(\tau)) \approx \sum_{i=0}^{m_v}\boldsymbol{\beta}_{l,i}F_i(\tau) \tag{58}$$

where $\boldsymbol{\beta}_{l,i}$ is the coefficient of Chebyshev polynomial with the same definition as Eq. (27), say

$$\boldsymbol{\beta}_{l,i} = \frac{2-\delta_{0i}}{m_v+1}\sum_{k=0}^{m_v}\mathbf{y}(\sigma_k,\mathbf{v}_l^e(\sigma_k))F_i(\sigma_k) \tag{59}$$

By analogy, it yields

$$\boldsymbol{\beta}_l = \left[ \beta_{l,0},\ \beta_{l,1},\ \cdots,\ \beta_{l,m_v} \right]^T$$

$$= \frac{2}{m_v + 1} \begin{bmatrix} \frac{1}{2}\mathbf{y}^T(\mathbf{v}_{l,0}^e)F_0(\sigma_0) + \cdots + \frac{1}{2}\mathbf{y}^T(\mathbf{v}_{l,m_v}^e)F_0(\sigma_{m_v}) \\ \mathbf{y}^T(\mathbf{v}_{l,0}^e)F_1(\sigma_0) + \cdots + \mathbf{y}^T(\mathbf{v}_{l,m_v}^e)F_1(\sigma_{m_v}) \\ \vdots \\ \mathbf{y}^T(\mathbf{v}_{l,0}^e)F_{m_v}(\sigma_0) + \cdots + \mathbf{y}^T(\mathbf{v}_{l,m_v}^e)F_{m_v}(\sigma_{m_v}) \end{bmatrix} \quad (60)$$

$$\triangleq \frac{2}{m_v + 1} \mathbf{Z}\mathbf{F}^T \mathbf{Y}$$

where $M$ is set to $m_v$ therein, and we denote $\mathbf{y}(\sigma_k, \mathbf{v}_l^e(\sigma_k)) \triangleq \mathbf{y}(\mathbf{v}_{l,k}^e)$ for the sake of representation. Regarding the calculation of matrix $\mathbf{Y}$, we have

$$\mathbf{Y} = \left[ \mathbf{y}(\mathbf{v}_{l,0}^e),\ \mathbf{y}(\mathbf{v}_{l,1}^e),\ \cdots,\ \mathbf{y}(\mathbf{v}_{l,m_v}^e) \right]^T$$

$$= \begin{bmatrix} \left( \mathbf{q}(\sigma_0) \circ \mathbf{f}^b(\sigma_0) \circ \mathbf{q}^*(\sigma_0) - 2\mathbf{w}^e \times \mathbf{v}_l^e(\sigma_0) + \mathbf{g}^e(\mathbf{p}_l^e(\sigma_0)) \right)^T \\ \left( \mathbf{q}(\sigma_1) \circ \mathbf{f}^b(\sigma_1) \circ \mathbf{q}^*(\sigma_1) - 2\mathbf{w}^e \times \mathbf{v}_l^e(\sigma_1) + \mathbf{g}^e(\mathbf{p}_l^e(\sigma_1)) \right)^T \\ \vdots \\ \left( \mathbf{q}(\sigma_{m_v}) \circ \mathbf{f}^b(\sigma_{m_v}) \circ \mathbf{q}^*(\sigma_{m_v}) - 2\mathbf{w}^e \times \mathbf{v}_l^e(\sigma_{m_v}) + \mathbf{g}^e(\mathbf{p}_l^e(\sigma_{m_v})) \right)^T \end{bmatrix}$$

$$= \begin{bmatrix} \left( \left[\mathbf{q}(\sigma_0)\right]^+ \left[\mathbf{f}^b(\sigma_0)\right]^+ \mathbf{q}^*(\sigma_0) - 2\left[\mathbf{w}^e\right]_\times \mathbf{v}_l^e(\sigma_0) + \mathbf{g}^e(\mathbf{p}_l^e(\sigma_0)) \right)^T \\ \left( \left[\mathbf{q}(\sigma_1)\right]^+ \left[\mathbf{f}^b(\sigma_1)\right]^+ \mathbf{q}^*(\sigma_1) - 2\left[\mathbf{w}^e\right]_\times \mathbf{v}_l^e(\sigma_1) + \mathbf{g}^e(\mathbf{p}_l^e(\sigma_1)) \right)^T \\ \vdots \\ \left( \left[\mathbf{q}(\sigma_{m_v})\right]^+ \left[\mathbf{f}^b(\sigma_{m_v})\right]^+ \mathbf{q}^*(\sigma_{m_v}) - 2\left[\mathbf{w}^e\right]_\times \mathbf{v}_l^e(\sigma_{m_v}) + \mathbf{g}^e(\mathbf{p}_l^e(\sigma_{m_v})) \right)^T \end{bmatrix} \quad (61)$$

$$\triangleq diag(\mathbf{Q}^*) diag(\mathbf{\Gamma}^{b+}) \mathbf{Q}^+ - 2 diag(\mathbf{V}_l^e) \mathbf{W}^{e\times} + \mathbf{G}^e(\mathbf{P}_l^e)$$

with the definition

$$\mathbf{Q}^* = \left[ \mathbf{q}^*(\sigma_0), \cdots, \mathbf{q}^*(\sigma_{m_v}) \right]^T$$

$$\mathbf{Q}^+ = \left[ [\overset{+}{\mathbf{q}(\sigma_0)}], \cdots, [\overset{+}{\mathbf{q}(\sigma_{m_v})}] \right]^T$$

$$\mathbf{\Gamma}^{b+} = \left[ [\overset{+}{\mathbf{f}^b(\sigma_0)}], \cdots, [\overset{+}{\mathbf{f}^b(\sigma_{m_v})}] \right]^T$$

$$\mathbf{V}_l^e = \left[ \mathbf{v}_l^e(\sigma_0), \cdots, \mathbf{v}_l^e(\sigma_{m_v}) \right]^T \tag{62}$$

$$\mathbf{W}^{e\times} = \left[ [\mathbf{w}^e]_\times, \cdots, [\mathbf{w}^e]_\times \right]^T$$

$$\mathbf{G}^e = \left[ \mathbf{g}^e(\mathbf{p}_l^e(\sigma_0)), \cdots, \mathbf{g}^e(\mathbf{p}_l^e(\sigma_{m_v})) \right]^T$$

$$\mathbf{P}_l^e = \left[ \mathbf{p}_l^e(\sigma_0), \cdots, \mathbf{p}_l^e(\sigma_{m_v}) \right]^T$$

where $\mathbf{q}(\sigma_k)$, $k=0,1,...,m_v$ are the attitude quaternion values at Chebyshev-roots points in the current iteration, which have already been obtained through the above attitude iterative computation. Besides, the specific force values $\mathbf{f}^b(\sigma_k)$, $k=0,1,...,m_v$ can be easily obtained from Eq. (16) or Eq. (18). $\mathbf{v}_l(\sigma_k)$, $k=0,1,...,m_v$ are the values of velocity at Chebyshev-roots points in the last iteration, and $[\mathbf{w}^e]_\times$ denotes the skew-symmetric matrix corresponding to the Earth rotation vector [26]. $\mathbf{g}^e(\mathbf{p}_l^e(\sigma_k))$, $k=0,1,...,m_v$ are the gravity vector values calculated by the gravity model in the local navigation frame but expressed in the ECEF frame, that is to say,

$$\mathbf{g}^e(\mathbf{p}_l^e(\sigma_k)) = \mathbf{C}_n^e \mathbf{g}^n(\mathbf{p}_l^n(\sigma_k)) = \mathbf{C}_n^e \mathbf{g}^n(ecef2lla(\mathbf{p}_l^e(\sigma_k))) \tag{63}$$

where we have

$$\mathbf{C}_n^e = \begin{bmatrix} -\sin L \cos \lambda & \cos L \cos \lambda & -\sin \lambda \\ -\sin L \sin \lambda & \cos L \sin \lambda & \cos \lambda \\ \cos L & \sin L & 0 \end{bmatrix} \tag{64}$$

$$\mathbf{g}^n = \begin{bmatrix} 0 & -g & 0 \end{bmatrix}^T \tag{65}$$

Both the attitude matrix described the transformation from the local navigation frame to the ECEF frame and the gravity model in the local navigation frame are functions of the curvilinear position, $\mathbf{p}_l^n = [\lambda \ L \ h]^T$, where $\lambda$ is the longitude, $L$ is the latitude, and $h$ is the height. The value of the curvilinear position is obtained by the Cartesian ECEF position in the last iteration, i.e., $\mathbf{p}_l^e$. To this end, the function $ecef2lla(\cdot)$ is used [30-32] in this paper. Besides, $g$ denotes the normal gravity in the vertical direction. The common WGS-84 gravity model is used following [33].

A big difference between the iNavFIter-M method and the iNavFIter method is that the former directly uses the position obtained in the last iteration for calculating the gravity vector, rather than approximating the gravity vector with Chebyshev polynomials in the iNavFIter. By so doing, significant computing burden has been saved. Meanwhile, the function values $\mathbf{y}(\mathbf{v}_{l,k}^e)$, $k=0,\cdots m_v$ allow parallel computation as well.

Furthermore, the velocity in the current iteration, namely, $\mathbf{v}_{l+1}^e(\tau)$ in Eq. (56), could also be approximated by a set of Chebyshev polynomials, say

$$\mathbf{v}_{l+1}^e(\tau) = \sum_{i=0}^{m_v} \mathbf{s}_{l+1,i} F_i(\tau) \tag{66}$$

Similarly, the following identity can be obtained in the light of Eqs. (56), (58), and (66)

$$\begin{aligned}
&\sum_{i=0}^{m_v} \mathbf{s}_{l+1,i} F_i(\tau) \\
&\approx \mathbf{v}^e(0) + \frac{t_N}{2} \int_{-1}^{\tau} \sum_{i=0}^{m_v} \boldsymbol{\beta}_{l,i} F_i(\tau) d\tau \\
&= \boldsymbol{\kappa}' + \frac{t_N}{2} \boldsymbol{\beta}_{l,0} F_1(\tau) + \frac{t_N}{2} \frac{\boldsymbol{\beta}_{l,1}}{4} F_2(\tau) + \frac{t_N}{2} \sum_{i=2}^{m_v} \frac{\boldsymbol{\beta}_{l,i}}{2} \left( \frac{F_{i+1}(\tau)}{i+1} - \frac{F_{i-1}(\tau)}{i-1} \right) \\
&= \boldsymbol{\kappa}' + \frac{t_N}{2} \left( \boldsymbol{\beta}_{l,0} - \frac{\boldsymbol{\beta}_{l,2}}{2} \right) F_1(\tau) + \frac{t_N}{4} \sum_{i=2}^{m_v-1} \frac{\boldsymbol{\beta}_{l,i-1} - \boldsymbol{\beta}_{l,i+1}}{i} F_i(\tau) + \frac{t_N}{4} \frac{\boldsymbol{\beta}_{l,m_v-1}}{m_v} F_{m_v}(\tau) + \frac{t_N}{4} \frac{\boldsymbol{\beta}_{l,m_v}}{m_v+1} F_{m_v+1}(\tau)
\end{aligned} \tag{67}$$

where $\boldsymbol{\kappa}'$ is defined as follows

$$\boldsymbol{\kappa}' = \mathbf{v}^e(0) + \frac{t_N}{2} \boldsymbol{\beta}_{l,0} - \frac{t_N}{8} \boldsymbol{\beta}_{l,1} + \frac{t_N}{2} \sum_{i=2}^{m_v} \boldsymbol{\beta}_{l,i} \frac{(-1)^{i+1}}{i^2-1} \tag{68}$$

Comparing the coefficients of Chebyshev polynomials from the 0th degree to the $m_v$th degree on both sides of Eq.(67), the following identity can be obtained

$$\begin{cases}
\mathbf{s}_{l+1,0} = \boldsymbol{\kappa}', \\
\mathbf{s}_{l+1,1} = \frac{t_N}{2} \left( \boldsymbol{\beta}_{l,0} - \frac{\boldsymbol{\beta}_{l,2}}{2} \right), \\
\mathbf{s}_{l+1,i} = \frac{t_N}{4} \frac{\boldsymbol{\beta}_{l,i-1} - \boldsymbol{\beta}_{l,i+1}}{i}, \quad i=2,\cdots,m_v-1 \\
\mathbf{s}_{l+1,m_v} = \frac{t_N}{4} \frac{\boldsymbol{\beta}_{l,m_v-1}}{m_v},
\end{cases} \tag{69}$$

Coefficients of all degrees can be put into a matrix, that is to say,

$$\mathbf{s}_{l+1} = \left[ \mathbf{s}_{l+1,0}, \ \mathbf{s}_{l+1,1}, \ \cdots, \ \mathbf{s}_{l+1,m_v} \right]^T \tag{70}$$

In summary,

$$\mathbf{s}_{l+1} = \begin{bmatrix} \mathbf{s}_{l+1,0}, & \mathbf{s}_{l+1,1}, & \mathbf{s}_{l+1,2}, & \cdots, & \mathbf{s}_{l+1,r}, & \cdots, & \mathbf{s}_{l+1,m_v} \end{bmatrix}^T$$

$$= \begin{bmatrix} \left(\mathbf{v}^e(0)\right)^T + \dfrac{t_N}{2}\boldsymbol{\beta}_{l,0}^T - \dfrac{t_N}{8}\boldsymbol{\beta}_{l,1}^T + \dfrac{t_N}{2}\sum_{j=2}^{m_v}\boldsymbol{\beta}_{l,j}^T \dfrac{(-1)^{j+1}}{j^2-1} \\ \dfrac{t_N}{2}\dfrac{1}{2\times 1}(2\boldsymbol{\beta}_{l,0}^T - \boldsymbol{\beta}_{l,2}^T) \\ \dfrac{t_N}{2}\dfrac{1}{2\times 2}(\boldsymbol{\beta}_{l,1}^T - \boldsymbol{\beta}_{l,3}^T) \\ \vdots \\ \dfrac{t_N}{2}\dfrac{1}{2\times r}(\boldsymbol{\beta}_{l,r-1}^T - \boldsymbol{\beta}_{l,r+1}^T) \\ \vdots \\ \dfrac{t_N}{2}\dfrac{\boldsymbol{\beta}_{l,m_v-1}^T}{2m_v} \end{bmatrix}$$

$$\triangleq \boldsymbol{\eta}_0 + \dfrac{t_N}{2}\mathbf{UD}\boldsymbol{\beta}_l = \boldsymbol{\eta}_0 + \dfrac{t_N}{2}\cdot\dfrac{2}{m_v+1}\mathbf{UDZF}^T\mathbf{Y}$$

$$= \boldsymbol{\eta}_0 + \dfrac{t_N}{m_v+1}\mathbf{C}_s\cdot\left(diag(\mathbf{Q}^*)diag(\boldsymbol{\Gamma}^{b+})\mathbf{Q}^+ - 2diag(\mathbf{V}_l^e)\mathbf{W}^{ex} + \mathbf{G}^e(\mathbf{P}_l^e)\right) \tag{71}$$

with the definition

$$\boldsymbol{\eta}_0 = \begin{bmatrix} \mathbf{v}^e(0), & 0, & \cdots, & 0 \end{bmatrix}^T \tag{72}$$

and the definition of matrix $\mathbf{C}_s$ is the same as Eq. (51), the maximum degree $M$ is set to $m_v$ herein.

In the light of Eq. (66), we have

$$\begin{aligned} \mathbf{V}_{l+1}^e &\triangleq \begin{bmatrix} \mathbf{v}_{l+1}^e(\sigma_0), & \cdots, & \mathbf{v}_{l+1}^e(\sigma_{m_v}) \end{bmatrix}^T \\ &= \begin{bmatrix} \mathbf{s}_{l+1,0}^T F_0(\sigma_0) + \mathbf{s}_{l+1,1}^T F_1(\sigma_0) + \cdots + \mathbf{s}_{l+1,m_v}^T F_{m_v}(\sigma_0) \\ \vdots \\ \mathbf{s}_{l+1,0}^T F_0(\sigma_{m_v}) + \mathbf{s}_{l+1,1}^T F_1(\sigma_{m_v}) + \cdots + \mathbf{s}_{l+1,m_v}^T F_{m_v}(\sigma_{m_v}) \end{bmatrix} \\ &= \begin{bmatrix} F_0(\sigma_0) & \cdots & F_{m_v}(\sigma_0) \\ \vdots & \vdots & \vdots \\ F_0(\sigma_{m_v}) & \cdots & F_{m_v}(\sigma_{m_v}) \end{bmatrix}\begin{bmatrix} \mathbf{s}_{l+1,0}^T \\ \vdots \\ \mathbf{s}_{l+1,m_v}^T \end{bmatrix} \\ &= \mathbf{F}\mathbf{s}_{l+1} \end{aligned} \tag{73}$$

where the definition of matrix $\mathbf{F}$ is the same as Eq. (40), but $M$ is set to $m_v$. Subsequently, these new obtained velocity values at Chebyshev-roots points will be used for the calculation of coefficients matrix in the next iteration.

Consequently, the Chebyshev coefficients describing the velocity function will be exactly computed until the accuracy criterion is met or the prescribed iteration number is reached. Similarly, the RMS error can also be chosen as the accuracy criterion, defined as

$$e_v = \sqrt{\sum_{i=0}^{m_v}(\mathbf{s}_{l+1,i}-\mathbf{s}_{l,i})^T(\mathbf{s}_{l+1,i}-\mathbf{s}_{l,i})\bigg/(m_v+1)} \tag{74}$$

## C. Position Computation

According to Eqs. (23) and (66), when the immediate velocity result is used, the iterative calculation of position is given by

$$\mathbf{p}_{l+1}^e(\tau) = \mathbf{p}^e(0) + \frac{t_N}{2}\int_{-1}^{\tau}\mathbf{v}_{l+1}^e(\tau)\,d\tau \approx \mathbf{p}^e(0) + \frac{t_N}{2}\int_{-1}^{\tau}\sum_{i=0}^{m_v}\mathbf{s}_{l+1,i}F_i(\tau)\,d\tau \tag{75}$$

Similarly, the position could also be approximated by Chebyshev polynomials with maximum degree of $m_p$,

$$\mathbf{p}_{l+1}^e(\tau) \approx \sum_{i=0}^{m_p}\mathbf{\rho}_{l+1,i}F_i(\tau) \tag{76}$$

Similar to Eq. (45), Chebyshev coefficients of velocity and position are related by

$$\begin{cases}
\mathbf{\rho}_{l+1,0} = \mathbf{\kappa}'', \\
\mathbf{\rho}_{l+1,1} = \dfrac{t_N}{2}\left(\mathbf{s}_{l+1,0} - \dfrac{\mathbf{s}_{l+1,2}}{2}\right), \\
\mathbf{\rho}_{l+1,i} = \dfrac{t_N}{4}\dfrac{\mathbf{s}_{l+1,i-1}-\mathbf{s}_{l+1,i+1}}{i}, \quad i=2,\cdots,m_p-1 \\
\mathbf{\rho}_{l+1,m_p} = \dfrac{t_N}{4}\dfrac{\mathbf{s}_{l+1,m_p-1}}{m_p},
\end{cases} \tag{77}$$

where $\mathbf{\kappa}''$ is defined as

$$\mathbf{\kappa}'' = \mathbf{p}^e(0) + \frac{t_N}{2}\mathbf{s}_{l,0} - \frac{t_N}{8}\mathbf{s}_{l,1} + \frac{t_N}{2}\sum_{i=2}^{m_p}\mathbf{s}_{l,i}\frac{(-1)^{i+1}}{i^2-1} \tag{78}$$

Correspondingly, let $\mathbf{\rho}_{l+1} = [\mathbf{\rho}_{l+1,0},\ \mathbf{\rho}_{l+1,1},\ \cdots,\ \mathbf{\rho}_{l+1,m_p}]^T$, the matrix form of the above identity is written as

$$\mathbf{\rho}_{l+1} = \varsigma_0 + \frac{t_N}{2}\mathbf{UDs}_{l+1} = \varsigma_0 + \frac{t_N}{2}\mathbf{C}_d\mathbf{s}_{l+1} \tag{79}$$

with the definition

$$\varsigma_0 = [\mathbf{p}^e(0),\ 0,\ \cdots,\ 0]^T \tag{80}$$

$$\mathbf{C}_d = \mathbf{U}\mathbf{D} \tag{81}$$

In the same way as Eq. (54) or Eq. (73), we have

$$\mathbf{P}_{l+1}^e = \mathbf{F}\boldsymbol{\rho}_{l+1} \tag{82}$$

with the definition

$$\mathbf{P}_{l+1}^e = \left[\mathbf{p}_{l+1}^e(\sigma_0), \mathbf{p}_{l+1}^e(\sigma_1), \cdots, \mathbf{p}_{l+1}^e(\sigma_{m_v})\right]^T \tag{83}$$

and $M$ is set to $m_p$ for position computation. Subsequently, the above obtained position values at Chebyshev-roots points will be used for calculating gravity vector matirx $\mathbf{G}^e$ in the next iteration.

It is obvious that the Chebyshev coefficients of position can be obtained without much effort once the updated Chebyshev coefficients of velocity are obtained. Therefore, the update of velocity and position can be performed together in one common iteration process. The RMS error is also chosen as accuracy criterion, defined as

$$\mathbf{e}_p = \sqrt{\sum_{i=0}^{m_p} \left(\boldsymbol{\rho}_{l+1,i} - \boldsymbol{\rho}_{l,i}\right)^T \left(\boldsymbol{\rho}_{l+1,i} - \boldsymbol{\rho}_{l,i}\right) \Big/ \left(m_p + 1\right)} \tag{84}$$

In conclusion, the proposed iNavFIter-M method is summarized in Fig. 2. It should be noted that the iterative process of velocity and position will not be started until the attitude iterative process is finished. It should be stressed that the matrices $\mathbf{C}_s$ and $\mathbf{C}_d$ are constant once the maximum degree of Chebyshev polynomials is prescribed, which can be computed prior to the iterative process for improving time efficiency. Furthermore, the proposed algorithm is inherently parallel since the values of function evaluated at Chebyshev-roots points are independent of each other.

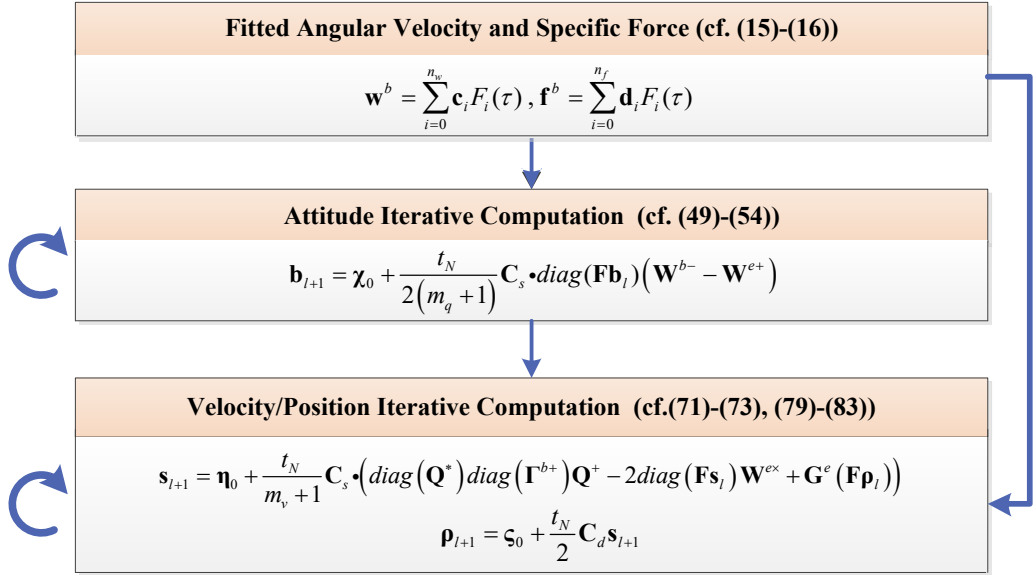

**Fig. 2** Flowchart of iNavFIter-M in the ECEF frame. Arrowed circles mean that two iterative computation processes continue until accurate criterion or prescribed iteration number is reached.

## IV. Numerical and FPGA Test Results

In this section, the flight test data sets in [18, 34] are used to evaluate the proposed iNavFIter-M algorithm against the state-of-the-art iNavFIter [18] as well as the typical two-sample algorithm [1]. The flight data sets are generated from specifically-designed trajectories with analytic ground true values for accuracy evaluation. Note that the simulation data are generated in the local navigation frame and then transformed into the Earth frame. The simulation scenario is the same as the iNavFIter work and the data sets are available in [35]. It is not for a realistic flight scenario but trying to mimic those cases that excite the noncommutativity error as far as possible.

In the numerical test, the sampling rate of gyroscope and accelerometer is 100 Hz, $N = 8$ samples of gyroscopes and accelerometers are used for fitting the angular velocity and the specific force. Consequently, the maximum degrees of fitted Chebyshev polynomials are set as $n_\omega = n_f = N - 1$. Besides, the maximum degrees of Chebyshev polynomials used to approximate attitude, velocity, and position are set as $m_q = m_v = m_p = N + 1$ to keep consistent with the iNavFIter. The maximum iteration number of all the iterative processes are uniformly set as $N + 1$. Additionally, the RMS error is adopted as the accuracy criterion with a threshold of $10^{-16}$.

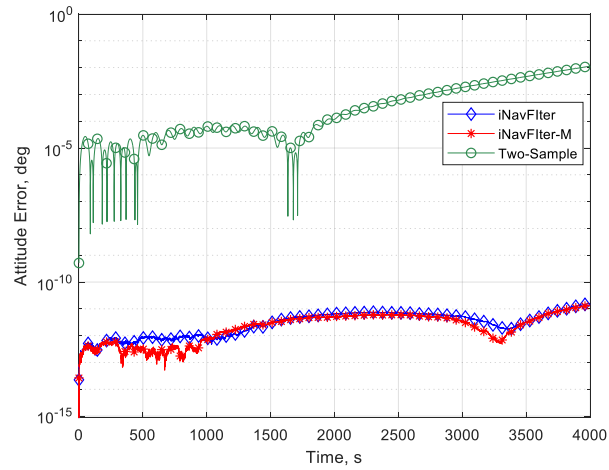

**Fig. 3 Principal angle errors of the iNavFIter algorithm (blue line), the iNavFIter-M algorithm (red line), and the typical two-sample algorithm (green line).**

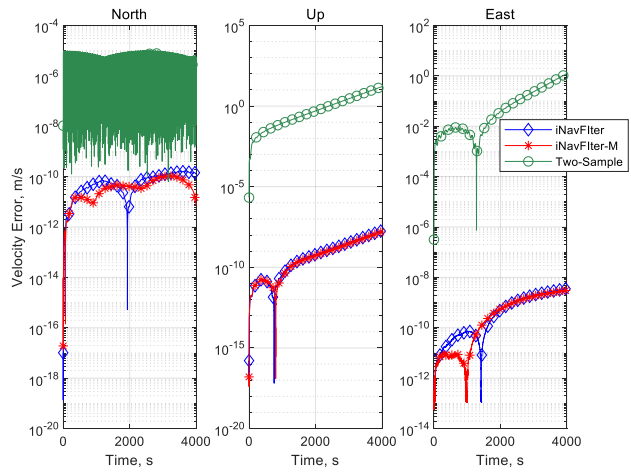

**Fig. 4 Velocity errors in the local navigation frame (north, up, east). Blue line denotes the results of iNavFIter, red line denotes the results of iNavFIter-M, and green line denotes the results of the typical two-sample algorithm.**

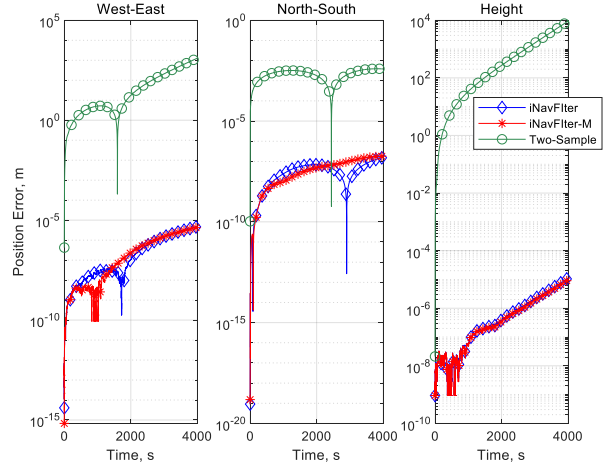

**Fig. 5 Position errors in the local navigation frame (west-east, north-south, height). Blue line denotes the results of iNavFIter, red line denotes the results of iNavFIter-M, and green line denotes the results of the typical two-sample algorithm.**

Figs. 3-5 compare the iNavFIter-M algorithm with both the iNavFIter and the typical two-sample algorithm in terms of computation errors of attitude, velocity, and position, respectively. The time duration of the simulation data is 4000 seconds, and the comparison is performed in the local navig]ation frame. The numerical results indicate that both iNavFIter-M and iNavFIter algorithms have a significant accuracy superiority of about 8-9 orders over the typical two-sample algorithm in all of attitude, velocity, and position accuracy. Note that iNavFIter-M is generally comparable to iNavFIter, marginally better in velocity and position. Specifically, their west-east position errors are listed in Table I. Table I also presents the average elapsed time of all algorithms across 10 runs on both the MATLAB platform and C++ platform without the parallel implementation. The numerical result indicates the proposed iNavFIter-M algorithm significantly reduces the computational burden of iNavFIter by over six times, even achieving comparable time efficiency with the typical two-sample algorithm on both MATLAB and C++ platforms. The significant efficiency gain is likely owed to the precomputation of constant matrix factors and the matrix-vector form operation. It is certain that the iNavFIter-M algorithm will be further accelerated when the parallel computation is implemented.

Table I    Performance comparison of iNavFIter-M, iNavFIter, and two-sample algorithm.

|  |  | iNavFIter-M | iNavFIter | Typical two-sample |
|---|---|---|---|---|
| Max West-East Errors | | 4μm | 4μm | 1260m |
| Elapsed Time | MATLAB | 49.2s | 295.3s | 53.1s |
|  | C++ | 9.3s | 59.4s | 5.9s |

Furthermore, our proposed iNavFIter-M algorithm is implemented and evaluated on the Field Programmable Gate Array (FPGA) platform. A photo of the FPGA platform is shown as Fig. 6. It consists of a Kinex-7 FPGA development board, a FL9031 ethernet module, and a JTAG downloader. The gyroscope and accelerometer sampling rates of test data are set as 100 Hz. The sensor data are fed to the FPGA development board by a laptop through the ethernet cable and the intermediate navigation result at each computing interval (0.08s) is also outputted from the same ethernet cable. The FPGA test results are shown as Fig. 7. It can be seen the iNavFIter-M algorithm also realizes high-precision computation on the FPGA platform, which further verifies the advantages of high accuracy and high efficiency of the proposed algorithm. At present, the iNavFIter-M algorithm can be run in real time on the FPGA platform to process inertial sensor data of sampling rate up to 1000 Hz.

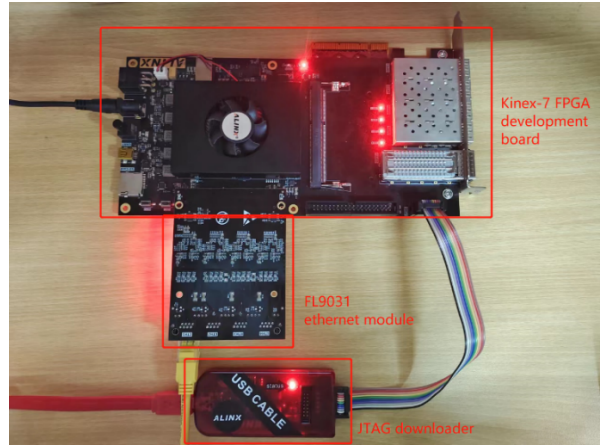

**Fig. 6  A shot of the FPGA platform while running the test.**

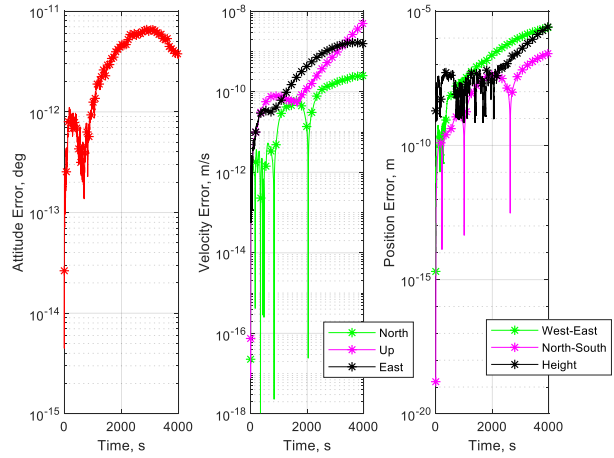

**Fig. 7 Errors of the principal angle, velocity and position in three axes in the local navigation frame of the iNavFIter-M algorithm evaluated on the FPGA platform.**

## V.  Conclusion

This paper proposes a matrix formulation of the functional iteration approach for inertial navigation computation, partly inspired by the orbit propagation field. Compared to our previous iNavFIter work, the proposed iNavFIter-M establishes the explicit connection of the Chebyshev polynomial coefficients between consecutive iterations, and notably spares the unnecessary computation burden spent on those polynomial terms higher than the truncation order. Numerical results have demonstrated that the iNavFIter-M not only maintains the accuracy superiority but improves the computational efficiency by about six times than the original iNavFIter, and is even comparable to the typical two-sample algorithm in running time. The iNavFIter-M algorithm has been evaluated and demonstrated on a FPGA board that can process in real time the inertial sensor data of up to 1000 Hz.

## Funding Sources

This work was supported in part by National Natural Science Foundation of China (62273228) and Shanghai Jiao Tong University Scientific and Technological Innovation Funds.

## Acknowledgments

Thanks to Dr. Qi Cai for the optimization of MATLAB code.